\documentclass[twocolumn]{article}
\usepackage{booktabs}
\usepackage{inputenc}
\usepackage[english]{babel}
\usepackage{amsmath}
\usepackage{amsfonts}
\usepackage{amssymb}
\usepackage{graphicx}
\usepackage{hyperref}
\usepackage{listings}
\usepackage{rotating}
\usepackage{authblk}
\usepackage[round]{natbib}
\usepackage{xcolor}

\providecommand{\keywords}[1]{\textbf{\textit{Keywords:}} #1}

\begin{document}

\title{When is the best time to learn? -- Evidence from an introductory statistics course}

\author[a]{Till Massing\thanks{Corresponding author. E-Mail:
till.massing@uni-due.de}}
\author[a]{Natalie Reckmann}
\author[a]{Alexander Blasberg}
\author[b]{Benjamin Otto\thanks{For more information regarding JACK contact this author.}}
\author[a]{Christoph Hanck}
\author[b]{Michael Goedicke}
\affil[a]{\footnotesize Faculty of Economics, University of
Duisburg-Essen, Universit{\"{a}}tsstr.~12, 45117 Essen, Germany}
\affil[b]{\footnotesize paluno - The Ruhr Institute for Software Technology,
University of Duisburg-Essen, Gerlingstr.~16, 45127 Essen, Germany}

\maketitle
\begin{abstract}
We analyze learning data of an e-assessment platform for an introductory mathematical statistics course, more specifically the time of the day when students learn. We propose statistical models to predict students' success and to describe their behavior with a special focus on the following aspects. First, we find that learning during daytime and not at nighttime is a relevant variable for predicting success in final exams. Second, we observe that good and very good students tend to learn in the afternoon, while some students who failed our course were more likely to study at night but not successfully so. Third, we discuss the average time spent on exercises. Regarding this, students who participated in an exam spent more time doing exercises than students who dropped the course before.
\end{abstract}

\keywords{Learning analytics, E-assessment, effective learning times} 

\section{Introduction}\label{sec:intro}
Modern university courses often use e-assessment systems. Especially when courses have a high number of participants e-learning tools are very useful to give students individual feedback. Courses with quantitative contents such as statistics and introductory mathematics are particularly suitable for e-assessment since exercises can also be designed as fill-in exercises -- which require students to submit a numeric answer -- instead of only multiple choice exercises. 
The e-assessment system JACK is a framework for delivering and grading complex exercises of various kinds \cite[]{Schwinning2017}. It was originally created to check programming exercises in Java \cite[]{Striewe2009}, but has been extended to several other exercise types such as multiple-choice and fill-in exercises \cite[]{Striewe2015a,Striewe2016,Schwinning2017}. JACK offers parameterizable content, meaning that exercises can contain different values each time an exercise is practiced. This means not only that different students get a different parameterization but, moreover, that the same student sees different numbers at each different time s/he selects the exercise. Hence, the exercise remains challenging until s/he understands the underlying concept to solve the exercise.

In addition to fill-in exercises, JACK allows to design exercises with dynamic programming content. For instance, JACK offers electronic Java or R -- the standard statistical programming language -- exercises. Programming exercises not only help to prepare students for modern statistical work, but also have been shown to be highly beneficial to foster their understanding of statistics \cite[]{Otto2017,Massing2018b}.

This study analyzes JACK data to more deeply understand students' learning behavior in an introductory mathematical statistics course. The high correlation between learning effort in the semester and the final grades is well documented \cite[and Section \ref{sec:related} provide examples]{Massing2018a}. Here, we aim to investigate additional aspects, namely the relevance of the daytime when students learn. More specifically, our research question is to analyze if it is possible to identify recommendations for students on when to best schedule their learning activities. To this end, we exploit granular data of the time stamps of student learning data made available by JACK. In order to do so, we use several statistical learning methods to study which factors in students' learning behavior are relevant to predict their success in the exam \cite[e.g., using tree-based methods as in][]{Hu2014}.  It turns out that daytime activity has a higher effectiveness than nighttime activity, which is in line with \cite{Gomes2011}.

Additionally, we find that good and very good students favor to learn in the afternoon, while some students who failed our course had insufficient learning behavior late at night. Moreover, we discuss the average time spent on exercises. Regarding this, students who participated at an exam spent more time on exercises than students who dropped the course before. This strengthens findings of \cite{Xiong2011}, who also found that response time is a relevant predictor for student performance. 

The remainder of this paper is organized as follows: Section \ref{sec:related} provides a brief overview of related work. Section \ref{sec:method} introduces the statistics course analyzed here and Section \ref{subsec:data} presents the available data and the models used. Section \ref{sec:analysis} discusses the empirical results. Section \ref{sec:conclusion} concludes.

\section{Related work}\label{sec:related}

The overall engagement of students is indisputably one of the main covariates of academic success. For the case of mathematical statistics this has been shown on several occasions. \cite{Sosa2011} show in a meta-study that the simultaneous usage of traditional classroom lectures and e-assessment has a positive effect on students' success. \cite{Massing2018a} substantiate the previous result by analyzing the learning activity on the e-assessment platform JACK. The study reveals that learning effort and success, measured by the total number of (correct) submissions on JACK over the course, positively affects the final grade in the exam. \cite{Otto2017} add additional R-programming exercises to the JACK framework and show that this exercise type helps to improve the general understanding of fundamental statistical concepts and thus ultimately yields better results in the final exam. 

Due to the empirically observed positive effect of a multitude of variables on academic performance, prediction of the latter has become possible. In this so-called branch of educational data mining various statistical learning methods are applied to educational data in order to predict student outcomes. Often, but not necessarily, this outcome is measured with a binary response of pass/fail in order to be able to provide an early-warning to students. \cite{James2013,Hastie2009,Meier2016} give a comprehensive overview of popular statistical learning methods used in the literature. For an overview of how to implement an early-warning-system see, e.g., \cite{binMat2013}. The literature has identified a number of important predictors. \cite{Gray2014} find evidence for the importance of socioeconomic and psychometric variables as well as pre-university grades, although \cite{Oskouei2014} show that, especially among the socioeconomic variables, the predictive capability can vary across countries. \cite{Baars2017} additionally identify post-admission variables like obtained credits, degree of exam participation and exam success rate to have an influence on students' success. \cite{Macfadyen2010,Wolff2013,Elbadrawy2015} analyze the learning activity on learning management systems and are able to accurately predict students' performance with appropriate variables. In a similar but more assessment-based fashion \cite{Huang2013,Burgos2018,Massing2018b} use activity in e-learning frameworks as well as the results of mid-term exams to predict students' success in the final exam. \cite{Asif2017} identify the performance in a small number of selected courses as a predictor for the academic achievement at the end of the study program. 

In contrast to previous studies which mostly rely on quantitative and qualitative learning activity measured by time-invariant variables, there is also a temporal dimension of engagement which has been studied from different perspectives throughout the academic literature. \cite{Hu2014} use time-dependent information provided by a learning management system to predict academic performance. \cite{Xiong2011} incorporate students' response time as an additional feature into a random forest to investigate the predictive capability for students' performance and find evidence that it can indeed improve prediction accuracy. \cite{Papamitsiou2014} and \cite{Papamitsiou2016} further elaborate on the latter by using more sophisticated techniques and are able to support the preceding result.

Only a few studies focus on the intraday engagement of students, that is, the actual daytime of learning, as a predictor for academic success. This topic is relevant as various studies show a significant influence of sleep quality and patterns on academic performance \cite[]{Ahrberg2012,Baert2015,Becsoluk2011,Eliasson2002}. Based on these insights \cite{Gomes2011} incorporate sleep variables into a prediction setting. With a stepwise regression approach they identify sleep frequency, night outings and sleep quality as among the most important predictors of academic success.\\\\

\section{Course Structure}\label{sec:method}

This section outlines the inital setup of the study. In particular, we sketch the structure of the analyzed course.

The e-assessment system JACK was used for a lecture and exercises course in mathematical statistics at the German university of Duisburg-Essen.
753 undergraduate first-year students started the course. The course is compulsory for several business and economics programs as well as in teachers' education.
Out of these 753 students, only 379 took an exam at the end of the course, while the others dropped the course in this term (Table \ref{tab:nostudents}). The course also introduces statistical programming skills using the statistics software R. In order to do so, the e-assessment system JACK offers programming exercises where the correctness of students' code is assessed, in addition to classical fill-in and multiple-choice exercises.

\begin{table}
	\center
	\resizebox{.45\textwidth}{!}{\begin{tabular}{lcc}
		\toprule
  		Students, who\ldots & counts & $\#$homework submissions \\
		\midrule
   	took course& 753 &163,444\\
		registered to an exam & 438 &152,232 \\
		participated at an exam & 379  &147,868 \\
		passed an exam & 186 &87,382\\
		\bottomrule
	\end{tabular}}
	\caption{Overview of the number of students registered to the course and the number of submissions on JACK.}
	\label{tab:nostudents}
\end{table}

The course consisted of a weekly 2-hours lecture, which introduced statistical concepts, and a 2-hours exercise class, which presented explanatory exercises and problems. Both classes were held classically in front of the auditorium. Due to the large number of students, these classes are limited in addressing students' different speeds of learning and individual questions. To overcome this issue and to encourage self-reliant learning, as well as to support students who had difficulties to attend classes, we offered all homework on JACK.

All in all, we offer 173 different exercises on JACK, of which 48 are designed as R-programming exercises and the remainder as fill-in or multiple-choice exercises. 
The individual learning success is supported by offering specific automated feedback and, furthermore, by optional hints. In case of additional questions which were neither covered by hints nor feedback, the students were able to ask questions in our moodle help-forum.

In order to further encourage students to learn continuously during the semester, and not only in the weeks prior to the exams, we offered five online tests using JACK. These tests lasted 40 minutes at fixed times in the evening. Four of the online tests contained fill-in or multiple-choice exercises only. The fifth online test contained R exercises exclusively. Participation only required a device with internet access, but no compulsory attendance at university. This summative assessment allowed students to assess their individual state of knowledge during the lecture period. It was not compulsory for students to participate at online tests in order to take the final exam at the end of the course.
Instead, we offered bonus points for the final exams to encourage participating at the tests (a maximum of 10 bonus points in total for fill-in online tests). The bonus points were only added to final exam points if at least 25 out of 60 exam points were reached, i.e., if students passed the exam without bonus.
The R online test was worth at most 2 bonus points which were awarded even if students achieved less than 25 points. The reason for this was to motivate students to focus on programming skills since \cite{Otto2017} and \cite{Massing2018b} show that this has a substantially (three times) higher impact on exam success than classical fill-in exercises. 

The final exams (3 in total) were also held electronically. While online tests during the semester could be solved at home with open books, the final exams were offered exclusively at university PC pools and supervised by academic staff. The exam consisted of R exercises ($\sim15\%$), short handwritten proofs ($\sim15\%$) and the remainder of fill-in exercises. Students can
only retake an exam if they failed or did not take the previous
ones (so that students can pass at most once), but can fail several
times.\footnote{Students obtain 6 “malus points” for each failed exam of which
they may collect at most 180 during their whole
bachelor program.} The last grade a student achieved in an exam will be denoted as the final grade. The corresponding exam will be denoted as the final exam.


\subsection{Data and Models}\label{subsec:data}
In this section we present the available database and the models used.
For each homework submission by a student on JACK we observe the exercise ID, the student ID, the number of points (on a scale from 0 to 100) and the time stamp with minute-long precision.

The response variables are given by the final exam success. We consider two possible responses. The first is a binary variable indicating whether a student passed (1) or did not pass (0) the course.
Second, we consider the final grade as a response. We have the following grading scheme: very good (``100''), good (``200''), satisfactory (``300''), sufficient to pass (``400'') and failed (``500'').
We assign ``600'' to students who took the course but did not participate at any of the exams. This is actually not a grade. However, this reflects the view that students who did not take any exam were even less prepared than students who failed the exams. Table \ref{tab:gradeoverview} reports an overview of final grades. We do not report grades for one specific exam date but the grade given at the end of the course.

\begin{table}
	\center
	\begin{tabular}{lc}
		\toprule
  		Grade (points) & counts \\
		\midrule
   	``100'' (48 -- 60) & 9 \\
		``200'' (40 -- 47) & 38  \\
		``300'' (31 -- 39) & 100   \\
		``400'' (25 -- 30) & 39 \\
		``500'' (0 -- 24) & 193 \\
		$\Sigma$ & 379 \\
		\hline
		failure rate &.508   \\
		``average grade'' &396.4 \\
		``standard deviation'' & 117.8\\
		\bottomrule
	\end{tabular}
	\caption{Overview of the distribution of grades, with the failure rate in the second--to--last line. The last line gives the ``average grade''.}
	\label{tab:gradeoverview}
\end{table}

\begin{figure}[ht]
	\centering
  \includegraphics[width=0.5\textwidth]{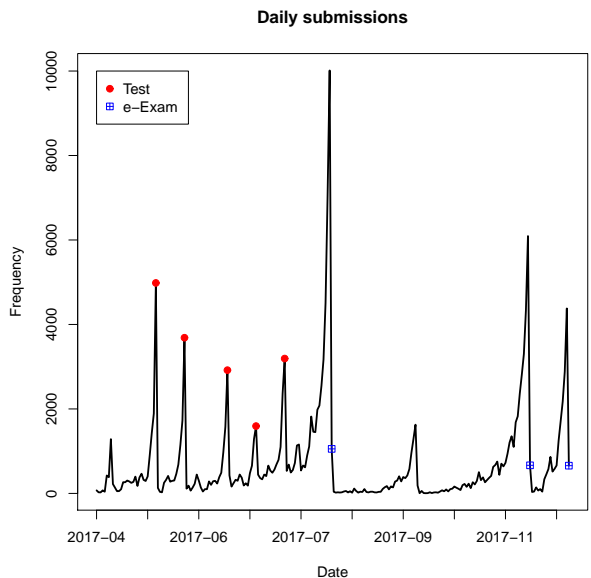}
	\caption{Number of daily submissions on JACK \textit{versus} the date. Exam and online test dates are highlighted with different points.}
	\label{fig:submissionvstime}
\end{figure}

JACK registered 163,444 submissions of homework exercises. See Table \ref{tab:nostudents} for how these submissions are distributed among students. Figure \ref{fig:submissionvstime} plots the number of daily submissions on JACK aggregated for all students. Characteristically, the number of submissions peaks shortly before a summative assessment such as an exam or an online test. This was also observed by \cite{Massing2018a}.

We compile the following information for each student $i$ from the raw data:
\begin{itemize}
\item the number of submissions (\# submissions in short),
\item the number of fully correct submissions (100 points),
\item the number of submissions in the morning from 8am to 12pm,
\item the number of submissions in the afternoon from 12pm to 4pm,
\item the number of submissions in the evening from 4pm to 8pm,
\item the number of submissions in the late evening from 8pm to 12am,
\item the number of submissions at night from 12am to 8am,
\item the median submission time (Subsection \ref{subsec:tageszeiten}).
\item The score, which is defined as follows:
let $t$ be a day during the semester. Then
\begin{equation*}
score_{it}:=\sum_{j=1}^n x_{ijt},
\end{equation*}
where $x_{ijt}$ is the number of points of the latest submission up to time $t$ of student $i$ in exercise $j$, $j=1,\ldots,n$. In other words, the score is the sum of
points of the last submissions to every exercise. This helps
tracking the learning progress for every student. In particular, we consider the final score, which is the score evaluated at the end of the term.
\item The frequency of submissions, i.e., the mean time between two following submissions at different days, measured in days.
\item The time until a student hands in the first submission from the beginning of the term, measured in days.
\item The time until a student hands in the last submission before his/her last exam, measured in days.
\item The number of days a student submitted solutions.
\item The average time spent per exercise measured in minutes (Subsection \ref{subsec:verweildauer}).
\end{itemize}

\begin{figure}[ht]
	\centering
  \includegraphics[width=0.5\textwidth]{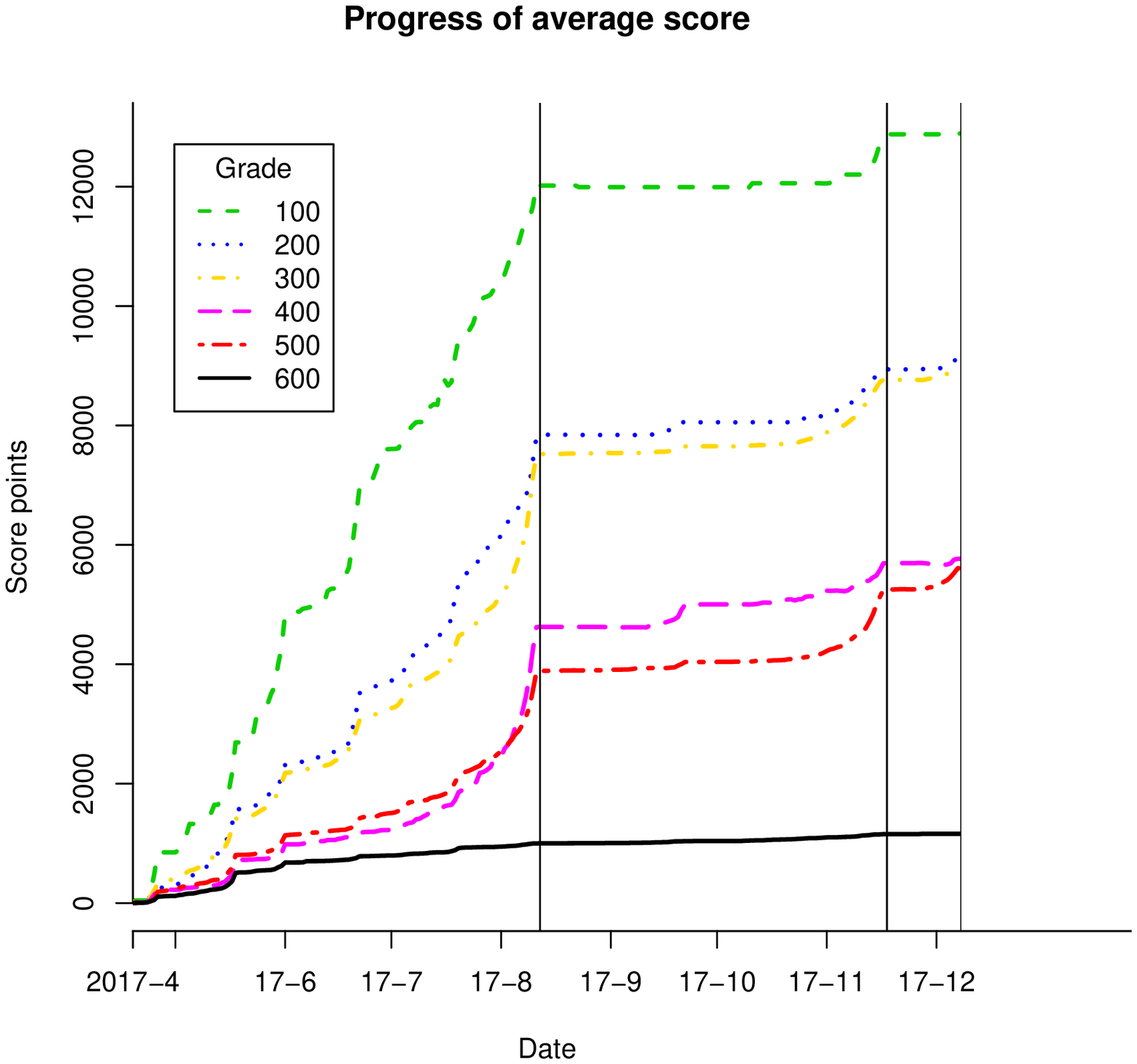}
	\caption{Average score (averaged over students with same grade) \textit{versus} the date. Thin vertical lines mark the exam dates.}
	\label{fig:score}
\end{figure}

Table \ref{tab:varsummary} reports summary statistics. Figure \ref{fig:score} plots the average score of students with different grades and of students who dropped the course. Evidently, good and very good students had a strong learning progress from the beginning of the semester on. Students with the sufficient pass grade ``400'' and students  who failed (``500'') start similarly weak but improve shortly before the exam. Students with ``400'', however, improve slightly more, which may be the reason that they pass the exam. On the other hand, they may just have been lucky in the exam. The students who dropped the course show very little progress on average.

\begin{table*}
	\center
	\begin{tabular}{lccccccc}
		\toprule
  		Variable name & Min & 1st Quartile &Median & Mean&3rd Quartile &Max &Sd\\
		\midrule
		\# submissions & 0&16&85&219.2&329&2726&296.8\\
		\# correct&0&7&32&94.7&126&1576&144.3\\
		\# morning&0&0&9&32.97&38&653&62.2\\
		\# afternoon&0&3&28&73.75&103&1273&111.7\\
		\# evening&0&5&32&71.56&105&819&97.9\\
		\# late ev&0&0&5&31.52&36&465&58.3\\
		\# night&0&0&0&9.432&5&396&32\\
		Median submission time&12:17am&2:16pm&3:56pm&3:37pm&7:19pm&11:31pm&2:52\\
		Final score&0&500&2372&4105&7333&15681&4100\\
		Frequency&1&5.25&9.556&50.433&26&234&84.8\\
		First submission&0&7&25.3&36.69&29.3&259.3&48.7\\
		Last submission&0&0.5&17.9&75.71&173.5&234&88.1\\
		\# days&0&2&9&15.47&25&90&16.6\\
		Ave Time spent&0&5.8&10.3&10.5&13.9&46.3&6.8\\
		\bottomrule
	\end{tabular}
	\caption{Overview of empirical quartiles mean and standard deviation for the considered covariates.}
	\label{tab:varsummary}
\end{table*}

We choose the following modeling approaches for the classification problem:\footnote{We also tried other modeling approaches than the ones stated here (e.g. neural networks, support vector machines, etc.). However their predictive performance proved not to be competitive.}
\begin{itemize}
\item \emph{Logistic Regression.} Logistic regression models the probability of an event given $p$ regressor variables $X =(X_1,\ldots,X_p)$ via
    \begin{equation}\label{logfunc}
    p(X) = \frac{\exp{\left(\beta_0 + \beta_1X_1+\ldots + \beta_pX_p\right)}}{1+ \exp{\left(\beta_0 + \beta_1X_1+\ldots + \beta_pX_p\right)}}.
    \end{equation}
     The idea is to regress the log-odds, $\log{\Big(\frac{p(X)}{1-p(X)}\Big)}$, on a linear combination of $X$. So equation \eqref{logfunc} can be rewritten as
    \begin{equation}
    \log{\left(\frac{p(X)}{1-p(X)}\right)} = \beta_0 + \beta_1X_1+\ldots,+\beta_p X_p.
    \end{equation}
    The unknown coefficients $\beta_0, \beta_1, \ldots, \beta_p$ are estimated based on the available data. We use maximum likelihood \cite[]{James2013}. We measure the variability of the estimates $\hat\beta_0, \hat\beta_1, \ldots, \hat\beta_p$ via the standard errors $SE{(\hat\beta_l)},~l=0,1,\ldots,p$, of the estimates. From this we obtain the $t$-statistics $t_l = \frac{\hat\beta_l}{SE(\hat\beta_l)}$.
    \cite{Kuhn2012} recommends to use the absolute value of the $t$-statistic of each non-constant regressor as importance measure for logistic regression.
		
    The above approach can easily be extended to ordered logistic regression in which we want to predict a variable with $k>2$ possible outcomes (multi-class-classifi-cation) \cite[]{Agresti2002}. We use binary logistic regression to predict the response ``student passed'', and ordered logistic regression to predict the grade.

\item \emph{Random Forests.} Tree-based methods can be used for two-class-classification as well as multi-class-classi-fication. 
     Single decision trees are very easy to interpret but have the drawback of having a high variance. To avoid this problem \cite{Breiman2001} proposed an algorithm for averaging decision trees to obtain a so-called random forest. The idea is to take $B$ bootstrap samples from the single training data set. Then, a tree is trained on every bootstrapped training data sample. Finally, the prediction is the majority vote, which is the most common occurring class over all $B$ predictions \cite[]{James2013}. Each of the single trees has a high variance but a low bias. Averaging over all trees reduces the variance. Another problem is that in each split of the trees, every variable in the predictor space is considered. If there is, for example, one very strong predictor it will be used in each tree for the first split. This leads to a high correlation between the trees. To avoid this problem, only a random sample of the $p$ predictors is used in each split to find an optimal split. The number of predictors in this random sample is usually set to $\sqrt{p}$. \cite{Breiman2001} also proposed to use the \textit{Mean Decrease Accuracy} as importance measure for the input variables. We build $500$ trees to grow the forest and try $3$ variables at each split.

\end{itemize}

\section{Empirical results}\label{sec:analysis}
This section analyzes students' learning behavior. We discuss which learning strategy turns out to predict students' exam success.

\subsection{Variable Importance}\label{subsec:varimp}

Our first analysis discusses which of the explanatory variables have a high predictive relevance. To model the target variable, i.e., passing the course or achieving a certain grade, we use the following set of variables as predictors in all of our models: \{\# correct, \# morning, \# afternoon, \# evening, \# late evening, \# night, score at day of first online test, score at day of third online test, final score, frequency, first submission, last submission, \# days, total time spent for exercises\}.
We dropped some variables like the total number of submissions to avoid high correlation between the predictor variables.

To compare the performance of the two models we use the accuracy, i.e., the rate of correctly classified observations. To avoid overfitting we use $3$-fold cross validation. Table \ref{tab:acc} contains the cross validation results for the two different models. We see that in both, two-class and multi-class classification, the random forest works best with an accuracy of $0.830$ or $0.73$ but logistic regression works well, too. In the full data set $75.3\%$ of the students do not pass the course, which leads to an accuracy of $0.753$ if we predict all students to not pass the course. Hence the random forest leads to an increase in accuracy of around $8$ percentage points which leads us to use the results of the random forest from now on.  

\begin{table}
	\center
\caption{Accuracy of both models}
	\resizebox{.45\textwidth}{!}{\begin{tabular}{lcc}
		\toprule
  		Model & Accuracy two-class & Accuracy multi-class \\
		\midrule
   	    Logistic Regression &  0.821 & 0.72  \\
		Random Forest       & 0.830 & 0.73  \\
		\bottomrule
	\end{tabular}}
	\label{tab:acc}
\end{table}

\begin{figure}[ht]
	\centering
  \includegraphics[width=0.45\textwidth]{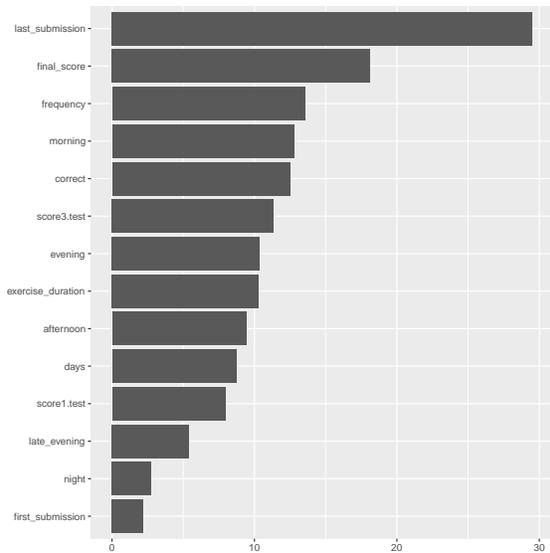}
	\caption{Variable importance measured in mean decrease accuracy for the two-class random forest.}
	\label{Imp1}
\end{figure}

We now investigate the variables which are chosen to build the single trees for the random forest. Figure \ref{Imp1} shows the importance of the variables used in the analysis.

\begin{figure}[ht]
	\centering
  \includegraphics[width=0.5\textwidth]{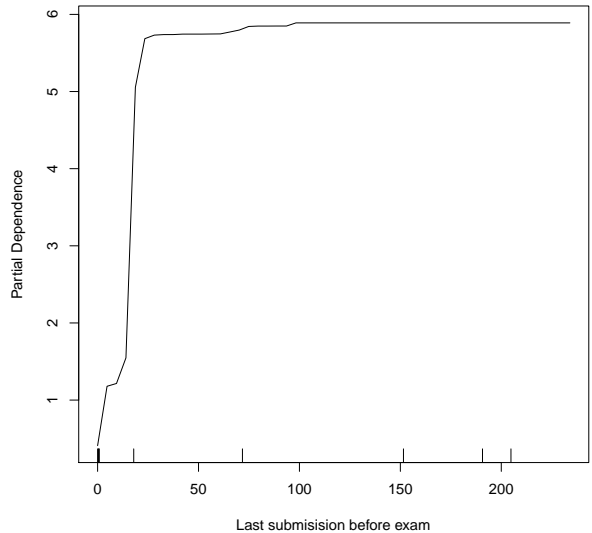}
	\caption{Partial dependence plot for the last submission variable before the exam for the multi-class random forest. Vertical ticks on the x-axis indicate deciles of the last submission variable.}
	\label{Partial}
\end{figure}

\begin{figure}[ht]
	\centering
  \includegraphics[width=0.5\textwidth]{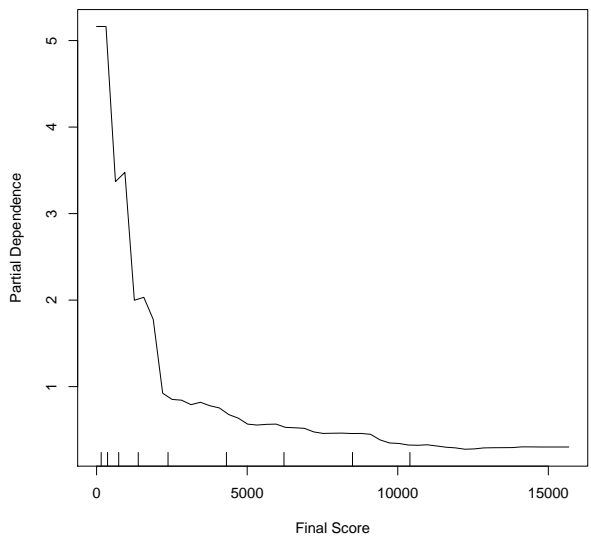}
	\caption{Partial dependence plot for the final score variable for the multi-class random forest.Vertical ticks on the x-axis indicate deciles of the final score variable.}
	\label{Partial2}
\end{figure}

We can see that the variable \textit{last\_submission}, i.e., the time until a student hands in the last submission before his/her last exam, measured in days, is by far the most important variable. Unfortunately, Figure \ref{Imp1} is silent on the direction of impact on the target variable. A solution to this problem is the partial dependence plot, which can help to understand how the log-odds of realizing the respective class depend on the input variables.\footnote{The y-axis shows $f(x) = \log p_k(x) - \frac{1}{K} \sum_{j=1}^K \log p_j(x)$, where $K$ is the number of classes, $k$ is the class of interest, and $p_j$ is the proportion of votes for class $j$.} A high positive value of the partial dependence for a given value of the predictor means that it is more likely to belong to the class of interest than to the other class, see \cite{Hastie2009}. Here the class of interest is \textit{not passing} the course. Figure \ref{Partial} shows the partial dependence plot for \textit{last\_submission}. We see that not passing the course is more likely for high values of \textit{last\_submission}. This means that students who learn until the day of the final exam unsurprisingly have a higher probability to pass the course than students who quit learning far before the exam. This is because 374 out of 753 students did not participate in an exam.
Most of these students did not learn until the exam but only made a few submissions at the beginning of the semester. Hence the variable \textit{last\_submission} has high values for these students. On the other hand most of the students who participated in the exams learned until shortly before the exam. This implies the high importance of the last submission.
Other important variables are the final score, the frequency of submissions and the number of submissions in the morning. Figure \ref{Partial2} shows the partial dependence plot of the final score. We see that a high final score leads to a low probability not to pass the course.\footnote{Note that in logistic regression the sign of the estimated coefficients tells the direction of the impact of a variable. These are mostly in line with the exemplary partial dependence plots.}

Furthermore, the importance of the variables in Figure \ref{Imp1} shows that the time of the first submission, the number of submissions at night and in the late evening and the first score in the term do not help for the predictive performance in the final exam. For the former and the latter this could be due to the fact that, at the beginning of the course, almost all students start to learn at the same level of knowledge, so there is no information that helps to decide between students passing or failing the final exam.

\begin{figure}[ht]
	\centering
  \includegraphics[width=0.45\textwidth]{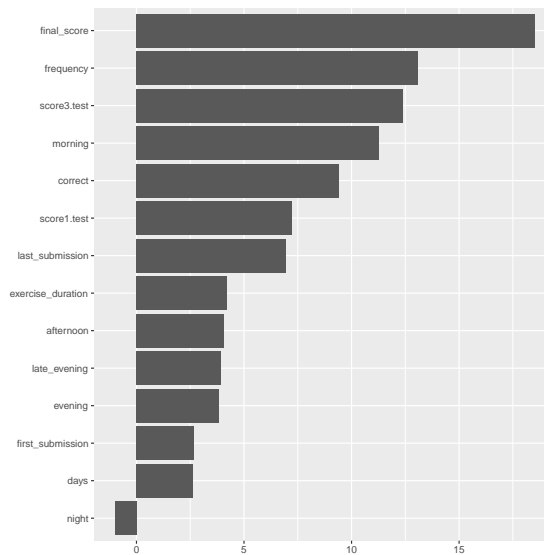}
	\caption{Variable importance measured in mean decrease accuracy for the two-class random forest without students who dropped the course.}
	\label{Imp2}
\end{figure}

\begin{figure}[ht]
	\centering
  \includegraphics[width=0.5\textwidth]{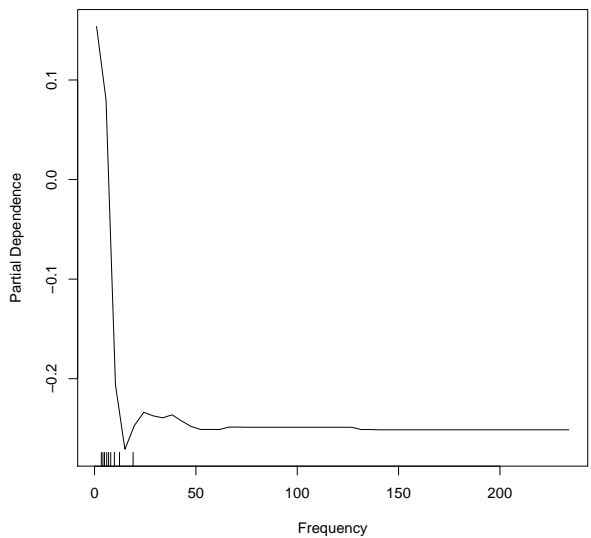}
	\caption{Partial dependence plot for the frequency variable for the two-class random forest without students who dropped the course. Vertical ticks on the x-axis indicate deciles of the frequency variable.}
	\label{Partial3}
\end{figure}

Since 374 out of 753 students did not participate in the exams we only focus on students who participated in an exam for the remainder of this subsection. This will obviously reduce the impact of the variable \textit{last\_submission}. We now estimate the corresponding binary classification random forest for pass \textit{vs.}~fail. Figure \ref{Imp2} shows its variable importance plot.\footnote{Note that a negative value for the mean decrease accuracy implies that randomly permuting the respective variable (\textit{ceteris paribus}) yields to a lower MSE of the random forest.} Now, final score and frequency, i.e., the mean time between two days of submissions measured in days, are the most important variables in the random forest model. For example, Figure \ref{Partial3} shows the partial dependence plot for the frequency. Small values of frequency make it more likely to pass the course. This means that students who learn regularly with only a few hours between their submissions have a higher probability to pass.

In case of multi-class classification, again, the time until a student hands in the last submission before his/her last exam is by far the most important variable in the model, for the same reasons as above. All other variables have low importance in this model. For reasons of brevity we shall now focus on the results of the binary model.

\subsection{Learning Times}\label{subsec:tageszeiten}
We now analyze more deeply at which time of the day good and less successful students prefer to learn. In order to investigate this we compute the median submission time for each student. We compare the median submission times for students who passed or did not pass in final exams.

\begin{figure}
	\includegraphics[width=0.5\textwidth]{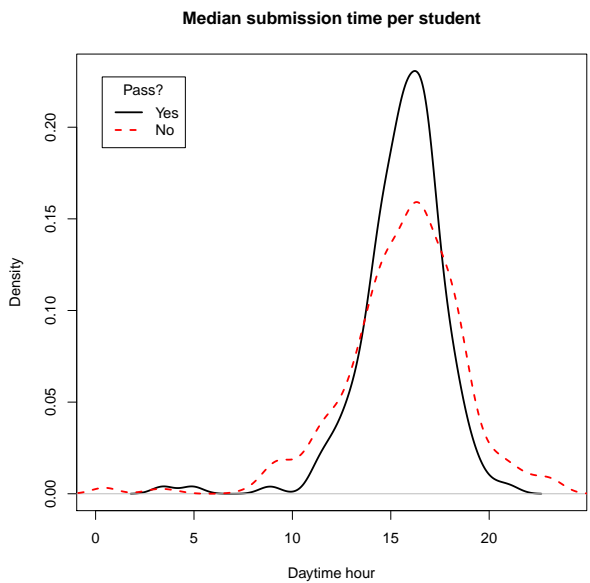}	
\caption{Kernel density estimates of the median daytime of submissions on JACK comparing students who passed with students who did not pass.}
\label{fig:mediansubpass}
\end{figure}

Figure \ref{fig:mediansubpass} shows kernel density plots for the median submission time for passing students in solid black and non-passing students in dashed red. There is a higher variance of median submission times for students who did not pass; students who passed prefer to learn in the afternoon. Weaker students tend to learn later. Moreover, quite a few non-passing students have median learning time in the morning. This is usually the time of the day when students should attend lecture and exercise classes.

\begin{figure}
	\includegraphics[width=0.5\textwidth]{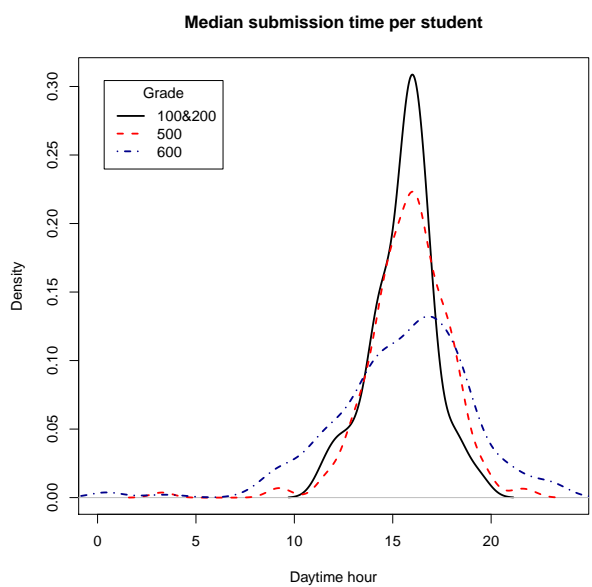}	
\caption{Kernel density estimates of the median daytime of submissions on JACK comparing students with different grades and with students who dropped the course.}
\label{fig:mediansubgrade}
\end{figure}

Figure \ref{fig:mediansubgrade} further supports this claim. We compare very good and good students with students who failed all exams and students who dropped the course. Evidently, good students prefer to submit exercises during daytime. The earliest median submission time of a good student is about 11:30am and the latest is 7:20pm.
Comparing these students with students who failed and, more visibly, who dropped shows that there are quite a few who study very late or very early. For example, the earliest is about 12:20am and the latest about 11:30pm.

This leads us to conclude that there are more non-passing students who have difficulties to learn in the afternoon. As stated in Section \ref{sec:related} lack of sleep caused by studying at night has a negative impact on students' performance.

Needless to say, the more important reason that poor students fail is mainly because they learn too little and not because of bad timing (cf.~Subsection \ref{subsec:varimp}). It also needs to be emphasized that unfavorable time management can also be due to a high amount of responsibilities not connected to their studies. Unfortunately, our data set does not allow us to distinguish between these aspects. A data set including both submission data for e-assessment and information on students' other daily activities is hard to collect.

\subsection{Submission duration}\label{subsec:verweildauer}

We now highlight another influential factor for success: we consider how long students work on a single submission, i.e., how much time they spent to solve an exercise. This analysis faces some challenges. First, we only observe the end and not the beginning of solving an exercise and hence do not have exact start and end times. We bypass this problem by measuring the time between two succeeding submissions. For example, if a student submits an exercise at 12pm and submits a second exercise (which might be the same as the first) at 12:15pm we consider 15 minutes as time spent for the second exercise. This means we do not observe duration of the first submission but of the following submissions. We omit duration times which are longer than two hours because students then likely took a break. This is also part of the second issue because we only monitor submission times in JACK and not whether students used this time to learn or whether they got distracted. We cannot rule out times of distractedness but still believe the following analysis offers interesting insights.

For each student we accumulate all duration times. These totals are of course higher for students who submitted more exercises than for students who submitted only a few. We thus divide total duration by the number of exercises submitted for each student.

\begin{figure}
	\includegraphics[width=0.5\textwidth]{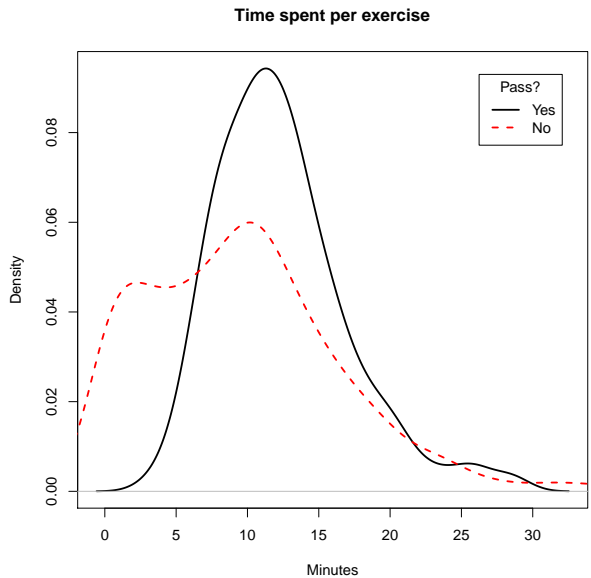}	
\caption{Kernel density estimates of the average time-spent for submissions comparing students who passed with students who did not pass.}
\label{fig:verweildauerpass}
\end{figure}

\begin{figure}
	\includegraphics[width=0.5\textwidth]{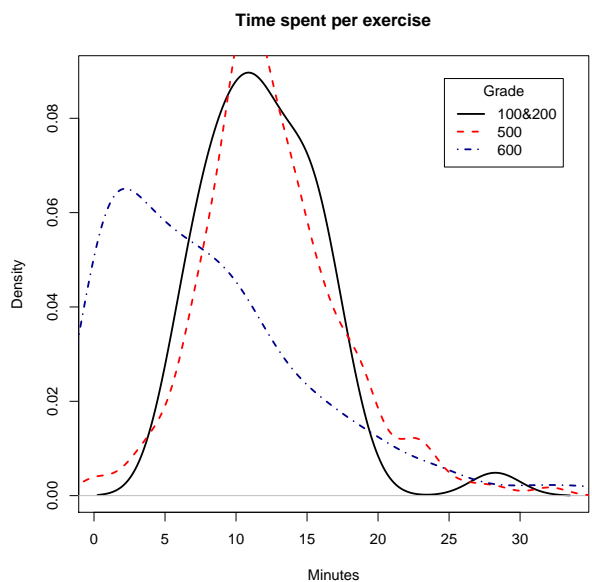}	
\caption{Kernel density estimates of the average time-spent for submissions comparing students with different grades and with students who dropped the course.}
\label{fig:verweildauernote}
\end{figure}

Figure \ref{fig:verweildauerpass} shows a kernel density estimate for the average time spent per submission of students who passed \textit{versus} students who did not pass. Evidently, there are many students who did not pass who invested little time for each exercise. Again, we next distinguish between students who failed an exam and students who dropped the course. Figure \ref{fig:verweildauernote} compares students who achieved the best or second best grade with students who failed and students who dropped the course. Interestingly, time spent per submission is similar for both good students and students who failed (The plot for mediocre students looks very similar, too). However, students who dropped the course invested perceptibly less time for each submission. Apparently, these students had too little motivation and/or time to participate in the course. They likely did not seriously attempt to solve the exercises.

\section{Conclusions and Discussion}\label{sec:conclusion}
This study analyzes when students should learn to be successful in a final exam. For this purpose, we analyzed data from the online learning platform JACK from an introductory mathematical statistics course in the summer term 2017. This data on students' submissions on JACK offered information about the daytime when a student submits a solution to an exercise.

We used logistic regression and random forests to predict the success of a student in the final exam and, also try to predict the final grade. An advantage of these methods is that they offer information about the importance of the variables used in the model. 
We analyze the variable importance obtained by the random forest.

The two most important variables in this model are the day between the last submission of an exercise and the exam as well as the score the students achieve when they study with JACK. We further identify the frequency with which the students work on JACK and the number of submissions between 8am and 12pm as important variables.
We identify good students to submit exercises during the daytime, while some students who quit the course or fail in the final exam learn very early in the morning or very late in the evening. Needless to say, the total amount of learning has a high impact on success. Additionally, we cannot rule out external factors (e.g. working during daytime) causing this effect rather than students who purposely did not study during daytime. Still we may conclude that students who did not pass the course study little during the afternoon.
Moreover the time a student spends on a single exercise is very short for students who dropped the course.

All in all, our results strengthen the evidence of the relevance of when and how often students learn. Suitable time management may increase students’ probability to pass a course like the one investigated here. That said, our study has some limitations which suggest potential for future work. First, we do not observe the reason why students learn at nighttime. However, these underlying reasons, such as childcare tasks or part-time work, may be the actual drivers of poor exam performance. Hence, the positive association between nighttime studying and poor performance may not so much be “bad” time management, but rather a conscious consequence of – at least some – students’ priorities that lead to lower academic success.

To estimate an actual causal effect of bad time management we would require such and possibly further information. One direction for future research is to conduct conditional surveys on JACK to obtain such relevant identifying information. A second limitation is that offline learning time, e.g., working with the lectures slides or textbooks is not covered in our setting since we cannot monitor such efforts electronically. However, we are confident that this aspect is not of dominant importance as high submission numbers suggest that students focus their efforts on JACK (which is also confirmed by informal student feedback). Third, as of now, JACK does not monitor “distractions” of students as, e.g., long time of passivity when working on an exercise (e.g., switching to other browser tabs, watching TV). Hence, a long measured duration of an attempt to solve an exercise may in fact not so much reflect sustained effort, but rather be due to a distraction. Future work using technical (such as mouse tracking) or survey-based approaches shall try to address these issues.

%

\section{Acknowledgments}
We thank all colleagues who contributed to the course ``Induktive
Statistik'' in the summer term 2017, especially Paul Navas Alban.
We thank Anna Janssen, Kim J.~Hermann, Janine Langerbein, Timo Rammert and Alexander Zyweck for excellent research assistance. \\
We gratefully acknowledge the valuable advice by two anonymous reviewers which helped to substantially improve this paper.\\
Part of the work on this project was funded by the German Federal
Ministry of Education and Research under grant numbers 01PL11075 and 01 JA 1610.

\bibliographystyle{plainnat}
\bibliography{bibliography}  

\end{document}